\documentclass[iop,revtex4]{emulateapj}
\pdfoutput=1
\usepackage{natbib}
\usepackage{afterpage}
\begin{document}

\title{The Bottom-Light Present Day Mass Function of the Peculiar Globular Cluster NGC 6535}
\author{Melissa Halford and Dennis Zaritsky}
\affil{Steward Observatory, University of Arizona, 933 North Cherry Avenue, Tucson, AZ 85721, USA} 
\email{mhalford@email.arizona.edu, dennis.zaritsky@gmail.com}

\begin{abstract}
Dynamical mass calculations have suggested that the Milky Way globular cluster NGC 6535 belongs to a population of clusters with high mass-to-light ratios, possibly due to a bottom-heavy stellar initial mass function. We use published Hubble Space Telescope data to measure the present day stellar mass function of this cluster within its half-light radius and instead find that it is bottom-light, exacerbating the discrepancy between the dynamical measurement and its known stellar content. The cluster's proximity to the Milky Way bulge and its relatively strong velocity anisotropy are both reasons to be suspicious of the dynamical mass measurement, but we find that neither straightforwardly explains the sense and magnitude of the discrepancy. Although there are alternative potential explanations for the high mass-to-light ratio, such as the presence of large numbers of stellar remnants or dark matter, we find this cluster to be sufficiently perplexing that we now exclude it from a discussion of possible variations in the initial mass function. Because this was the sole known old, Milky Way cluster in the population of high dynamical mass-to-light ratio clusters, some possible explanations for the difference in cluster properties are again open for consideration.
\newline
\end{abstract}


\section{INTRODUCTION}
\label{Introduction}

The stellar initial mass function (IMF) is key to understanding the details of star formation and implicit in many measurements made in extragalactic astronomy. Our particular interest lies in the interplay between the behavior of the low-mass end of the IMF and the resulting total mass of a stellar population. Several recent studies using a variety of techniques have suggested that variations in the low-mass end of the IMF exist among different stellar populations (e.g. \citealt{Cappellari}, \citealt{CvD}, \citealt{Geha}, \citealt{Spiniello}), but the indirectness of some of these methods, large systematic uncertainties, and the importance of these results to much of extragalactic astronomy make further investigation necessary.

Studies of stellar clusters complement those of galaxies because, unlike the stars in galaxies, we can safely presume that the stars in clusters have similar ages and metallicities, and because for some clusters we can resolve individual low luminosity stars. Using new, more precise velocity dispersion measurements for a set of stellar clusters in the Milky Way and its satellite galaxies, \citet{Z12,Z13,Z14} determined cluster dynamical masses, calculated stellar population mass-to-light ratios ($\Upsilon_*$), and used stellar evolution models to evaluate the corresponding mass-to-light ratios for an age of 10 Gyr, ($\Upsilon_{*,10}$). In doing so, they identified a low $\Upsilon_{*,10}$ population consisting mainly of old (age $>10$ Gyr) clusters of a wide range of metallicities ($-2.1<[$Fe/H$]<0$) and a high $\Upsilon_{*,10}$ population consisting mainly of young (age $<10$ Gyr), more metal-rich ($-1<[$Fe/H$]<0$) clusters. Zaritsky et al. noted that the differences in $\Upsilon_{*,10}$ may correspond to the same scale of IMF variations hypothesized for galaxies, where the low $\Upsilon_{*,10}$ clusters have an IMF that is consistent with that measured in our Galaxy \citep[and references therein]{Bastian} including modest dynamical evolution of the clusters, and the high $\Upsilon_{*,10}$ clusters have a bottom-heavy IMF, consistent with what is claimed to be the case in early type galaxies \citep{Cappellari,CvD,Spiniello}.

Studies of stellar clusters face some unique challenges. Large binary fractions, where the binary orbital velocities are larger than the internal cluster velocity dispersion, discrepancies between the phase space distribution function of actual clusters and that assumed in the dynamical models used to calibrate and test the mass estimation technique, internal dynamical relaxation, and external tidal influences can all lead to inaccurate cluster mass measurements. Beyond problems with the mass estimation, problems with the evolutionary models can cause errors in the calculated values of $\Upsilon_{*,10}$ that vary with age or metallicity. Lastly, even if those issues are minor, large numbers of stellar remnants or dark matter may cause differences in $\Upsilon_{*,10}$ that are unrelated to IMF behavior. Therefore, independent and direct confirmation of the low-mass end of the stellar mass function is absolutely necessary in cases where deviations from the norm are suspected. For sufficiently nearby clusters, the most direct method to determine the stellar mass function is to count stars.

Among the current sample of high $\Upsilon_{*,10}$ clusters, only one, NGC 6535, is located in the Milky Way, making it sufficiently close for us to resolve stars well into the subsolar mass regime. This cluster is also unique within the sample of high $\Upsilon_{*,10}$ clusters because it is the only old cluster (age $>10$ Gyr). As such, it provides the only evidence to date that the $\Upsilon_{*,10}$ differences are not solely due to age, host galaxy, or systematic errors in the stellar evolution models.

To determine whether the high dynamical mass-to-light ratio of NGC 6535 is due to an excess of low-mass stars, we measure the PDMF of this cluster and compare it to those of other clusters in \S\ref{PDMF}. We discuss the results and implications in \S\ref{Discussion}.

\section{PRESENT DAY MASS FUNCTION}
\label{PDMF}

\subsection{NGC 6535 and  Other Clusters of Similar Age and Metallicity}
\label{Comparison}

To construct a mass function down to sufficiently low stellar masses, we require deep, high resolution imaging of NGC 6535. The ACS Survey of Galactic Globular Clusters was a Hubble Space Telescope program (GO-10775)  that targeted Milky Way globular clusters for imaging with the Wide Field Channel (WFC) of the Advanced Camera for Surveys \citep{Survey}. With a few exceptions that we do not consider here, each cluster was observed for one orbit in F606W and one orbit in F814W using one short exposure and four to five long exposures in each filter. The WFC images have a resolution of 50 mas pixel$^{-1}$ and cover 1.7$^\prime \times 3.4^\prime$ on each of the two CCDs. The long exposures were dithered so that stars located in the gap between the instrument's CCDs in one exposure would be on a detector in the other images. \citet{Catalogs} provide photometry and catalogs of artificial stars that can be used for completeness corrections. Because of possible systematic errors in the conversion of the photometry to a PDMF, we study a comparison set of clusters of similar age and metallicity to NGC 6535 from the published survey so that we can do a mostly model-free relative comparison. Quantitatively, we include clusters with ages that differ by less than 5\% from that of NGC 6535, which, based on the normalization in \citet{Ages}, corresponds to ages that differ by less than 0.64 Gyr. From the set of clusters in the ACS survey that satisfy this criterion, we exclude NGC 6715 because of signs of contamination in its CMD, NGC 2808 because of unacceptable scatter in the CMD, NGC 1851 because of low completeness along most of the main sequence, and NGC 362 because it lacks artificial star data in the archive. These considerations leave us with the comparison clusters listed in Table \ref{Parameters}.

\afterpage{%
\begin{turnpage}
\begin{deluxetable*}{lccccccccccc}
\tabletypesize{\scriptsize}
\tablewidth{0pt}
\tablecaption{Parameters of NGC 6535 and Comparison Clusters}
\tablehead{
\colhead{Cluster} &
\colhead{[Fe/H]} &
\colhead{Age} &
\colhead{$r_e$} &
\colhead{$r_{max}$} &
\colhead{$m_{F606W,max}$} &
\colhead{$D_{I}$} &
\colhead{$D_{H}$} &
\colhead{$D_{E}$} &
\colhead{$E(B-V)_{I}$} &
\colhead{$E(B-V)_{H}$} &
\colhead{$E(B-V)_{E}$} \\ &&
\colhead{(Gyr)}&\colhead{$(\ \prime \ )$}&\colhead{$(\ \prime \ )$}&&\colhead{(kpc)}&\colhead{(kpc)}&\colhead{(kpc)}&&& \\
}
\startdata
NGC 288 & $-1.32$ & 10.6 & 2.23 & 1.63 & 25.5 & 9.4 & 8.9 & 9.9 & 0.05 & 0.03 & 0.03 \\
NGC 1261 & $-1.27$ & 10.2 & 0.68 & 0.68 & 22.2 & 17.4 & 16.3 & 17.2 & 0.02 & 0.01 & 0.01 \\
NGC 3201 & $-1.59$ & 10.2 & 3.10 & 1.67 & 25.5 & 4.8 & 4.9 & 4.9 & 0.33 & 0.24 & 0.28 \\
NGC 5904 & $-1.29$ & 10.6 & 1.77 & 1.58 & 21.2 & 7.8 & 7.5 & 7.8 & 0.04 & 0.03 & 0.04 \\
NGC 6535 & $-1.79$ & 10.5 & 0.85 & 0.85 & 25.1 & 6.5 & 6.8 & 6.5 & 0.51 & 0.34 & 0.45 \\
Arp 2 & $-1.75$ & 10.9 & 1.77 & 1.67 & 25.5 & 30.4 & 28.6 & 30.6 & 0.13 & 0.10 & 0.10 \\
NGC 6934 & $-1.47$ & 11.1 & 0.69 & 0.69 & 22.2 & 16.4 & 15.6 & 16.4 & 0.11 & 0.10 & 0.11 \\
NGC 6981 & $-1.42$ & 10.9 & 0.93 & 0.93 & 24.1 & 17.5 & 17.0 & 17.8 & 0.06 & 0.05 & 0.05 \\
\enddata
\tablecomments{Metallicities and half-light radii from \protect\citet[2010 edition]{Harris}. Ages from \protect\citet{Ages}. See text for definitions of the various distance and extinction values.}
\label{Parameters}
\end{deluxetable*}
\end{turnpage}
}

We construct completeness-corrected F606W vs. F606W-F814W color-magnitude diagrams (CMDs) for NGC 6535 and the comparison clusters. As suggested in \citet{Catalogs}, when calculating completeness we only consider an artificial star ``recovered'' if the detected star deviates from the input star by less than 0.5 pixels in position and 0.75 in instrumental magnitude in both bands. These criteria reduce the probability that a real star in the image is counted as a recovered artificial star. To account for completeness variations that are both position- and flux-dependent, we calculate completeness corrections in one magnitude wide bins in F606W over annuli that each cover 100 pixels in radius from the center of the cluster. To avoid uncertainties due to large corrections, we exclude from our analysis bins where the completeness is $<0.5$. The limiting magnitude for each cluster is set by this completeness limit or $m_{F606W}=25.5$, whichever is lower. We present and discuss the stellar mass function out to $r_{max}$, the smaller of the cluster's half-light radius or the size of the image. We use the half-light radius because it corresponds to the radius of the velocity dispersion measurements used to calculate mass-to-light ratios in \citet{Z12,Z13,Z14}. We present the various resulting limits in Table \ref{Parameters}.

We calculate the correspondence between stellar magnitudes, colors, and masses using isochrones from the Dartmouth Stellar Evolution Database \citep{Isochrones}. We use metallicity, distance, and extinction values from the compilation of cluster properties by \citet[2010 edition]{Harris} and ages from \citet{Ages}. The comparisons we present in this part of our analysis are among clusters of similar relative ages and so independent of the more uncertain absolute age. To optimize the isochrone fit, we fix metallicity and age to the literature values while we vary distance and extinction. The full set of adopted parameters is listed in Table \ref{Parameters} and described in more detail below.

To exclude contaminating field stars, we define a ridgeline in each CMD and remove stars that are far from this ridgeline in color. We define the position of the ridgeline at the magnitude of each star using the median F606W-F814W color of stars in a 0.1 magnitude bin centered on that magnitude. We exclude stars more than $3\sigma$ from the ridgeline from subsequent analysis, where $\sigma$ is the standard deviation of the F606W-F814W color in the same bin, calculated separately in the blue and red directions.

\begin{figure}[t]
\centering
\includegraphics[scale=0.45]{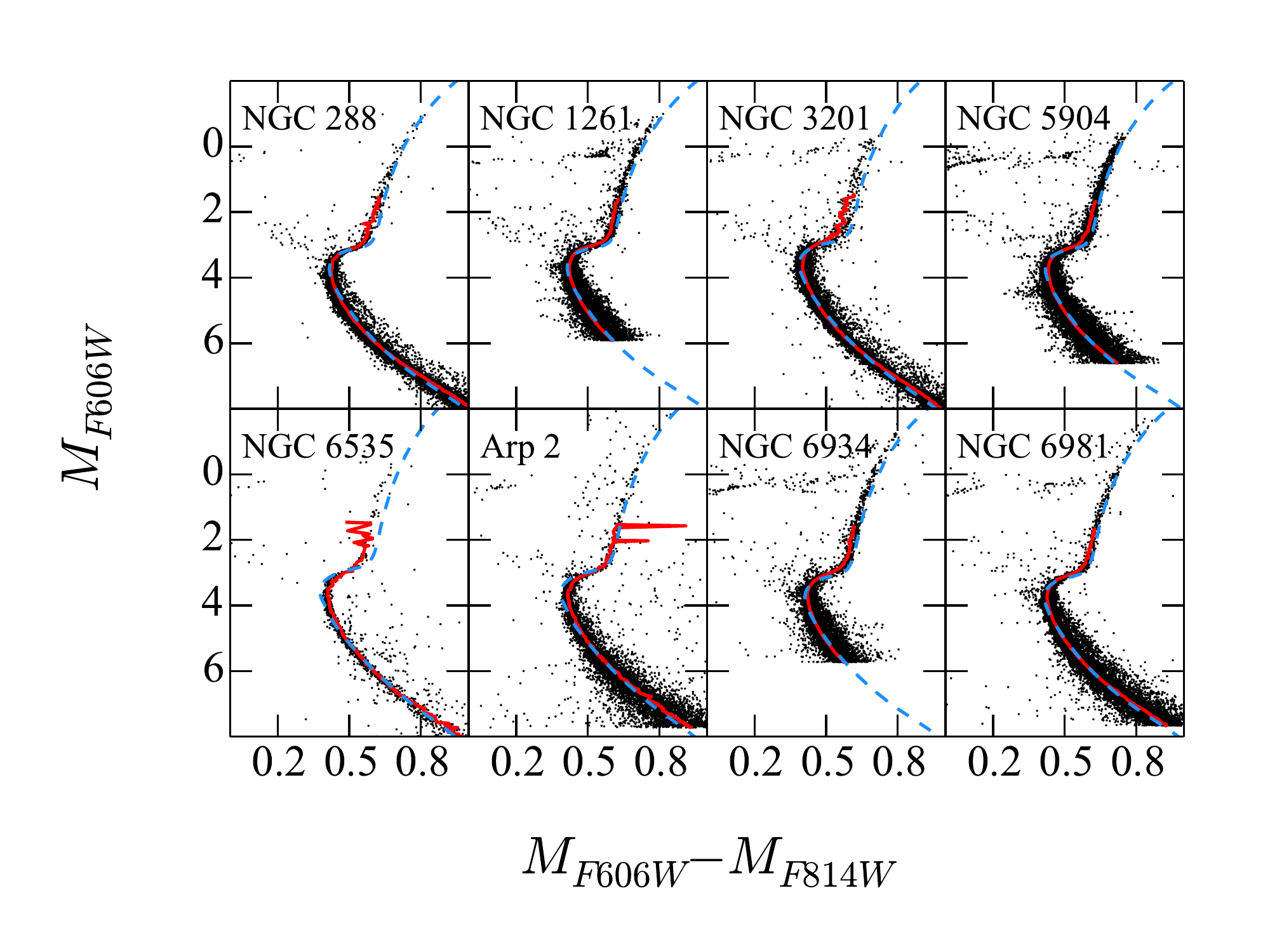}
\caption{Color-magnitude diagrams of NGC 6535 and other clusters of similar age and metallicity. The ridgeline (solid red line) and isochrone (dashed blue line) are also plotted for each cluster. The distances and extinctions are calculated by a least squares fit with metallicity, age, and [$\alpha$/Fe] fixed to literature values. See text for details.}
\label{CMD}
\end{figure}

\begin{figure}[t]
\centering
\includegraphics[scale=0.45]{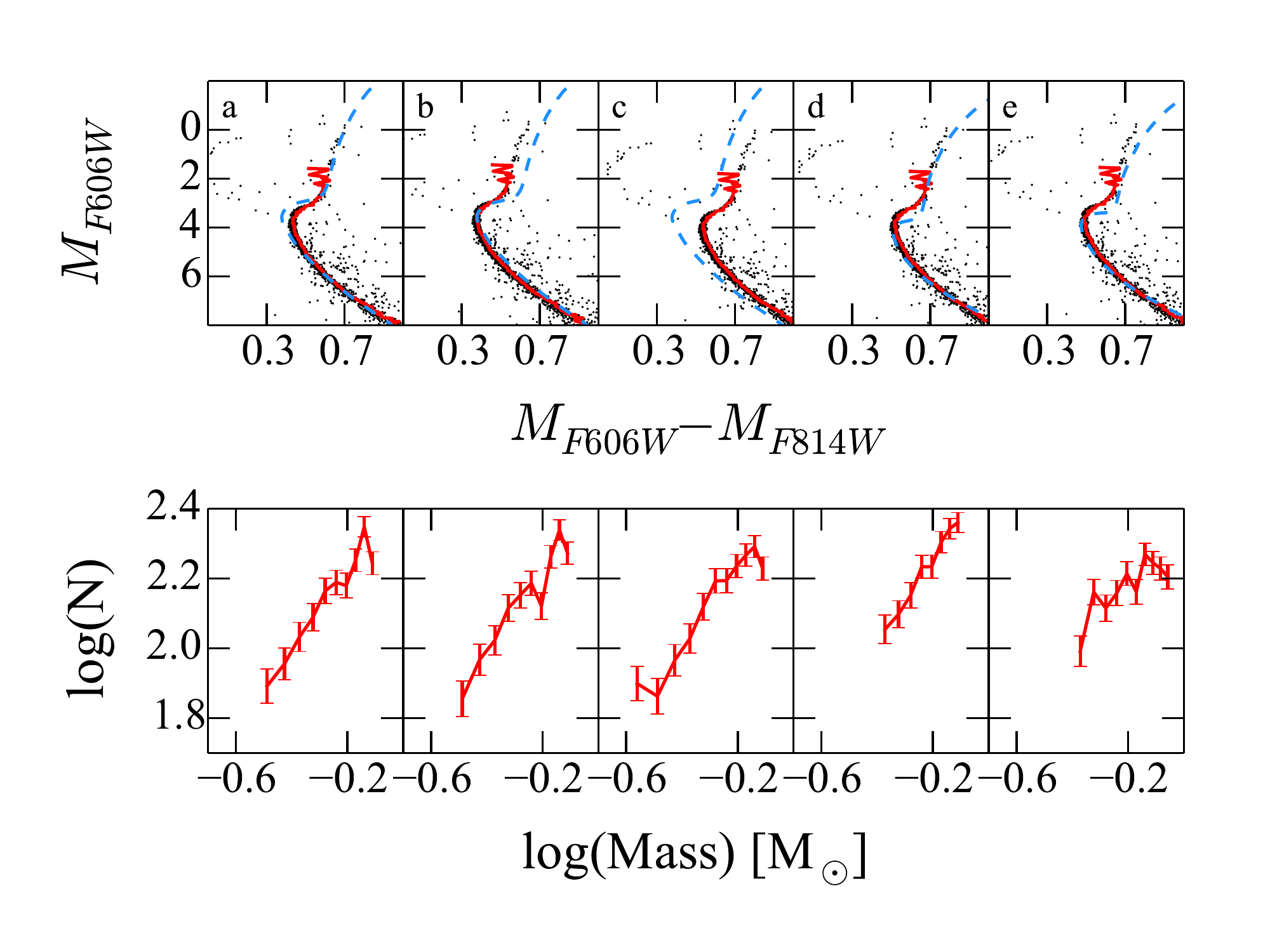}
\caption{Color-magnitude diagrams (top row) and corresponding PDMFs (bottom row) of NGC 6535 with different isochrone fits. The ridgeline (solid red line) and isochrone (dashed blue line) are plotted in each color-magnitude diagram. The isochrones used are a) an empirical fit of the isochrone to that of NGC 5904, b) the best fit distance and extinction keeping metallicity, age, and [$\alpha$/Fe] fixed to literature values, c) all parameters fixed to literature values, d) the best fit distance, extinction and metallicity with age and [$\alpha$/Fe] fixed, and e) the best fit with all parameters free to vary. Independent of the specifics of the isochrone fits, the PDMF declines with decreasing stellar mass. See text for details.}
\label{6535}
\end{figure}

To explore the sensitivity of our results to parameter choices, we calculate the mass functions using a variety of plausible parameters. First we use the model isochrones to determine the distance, $D_{I}$, and extinction, $E(B-V)_{I}$, using a least squares fit of the isochrone to the ridgeline within the magnitude range $m_{F606W,MSTO}-2<m_{F606W}<m_{F606W,max}$ where $m_{F606W,MSTO}$ is the F606W magnitude of the main sequence turnoff (MSTO) from \citet{Ages} and $m_{F606W,max}$ is the magnitude limit set by the 0.5 completeness criterion and the limit at $m_{F606W}=25.5$. Second, we use the published values of the distances, $D_{H}$, and extinctions, $E(B-V)_{H}$, from \citet[2010 edition]{Harris}. Finally, we define an empirical distance, $D_{E}$, and extinction, $E(B-V)_{E}$, by fitting the ridgeline of each cluster to the ridgeline of the most populated cluster in the sample (NGC 5904) and using the least squares fit to the isochrone for NGC 5904. To calculate the reddening, we use $A_\lambda/E(B-V)$ values from \citet{Extinction}. We use [$\alpha$/Fe]=0.2 for the isochrones. Figure \ref{CMD} contains the CMDs for the clusters using absolute magnitudes determined by $D_{I}$ and $E(B-V)_{I}$ with the ridgelines and isochrones overplotted. It is clear from the figure that the isochrone for NGC 6535 does not fit very well along the giant branch, but we performed our analysis using a variety of isochrones and were unable to find one that agreed well with the data. Fits that are slightly better have parameters that are far from the accepted values for this cluster. We show in Figure \ref{6535} the CMDs and corresponding PDMFs for each of the isochrone options we explored for this cluster. The results are sufficiently similar for all of the options that none of our conclusions are affected by the choices of these parameters, within the ranges explored.  The results that follow use $D_{I}$ and $E(B-V)_{I}$ with metallicity, age, and [$\alpha$/Fe] fixed to literature values for all clusters.


\begin{figure}[t!]
\centering
\includegraphics[scale=0.45]{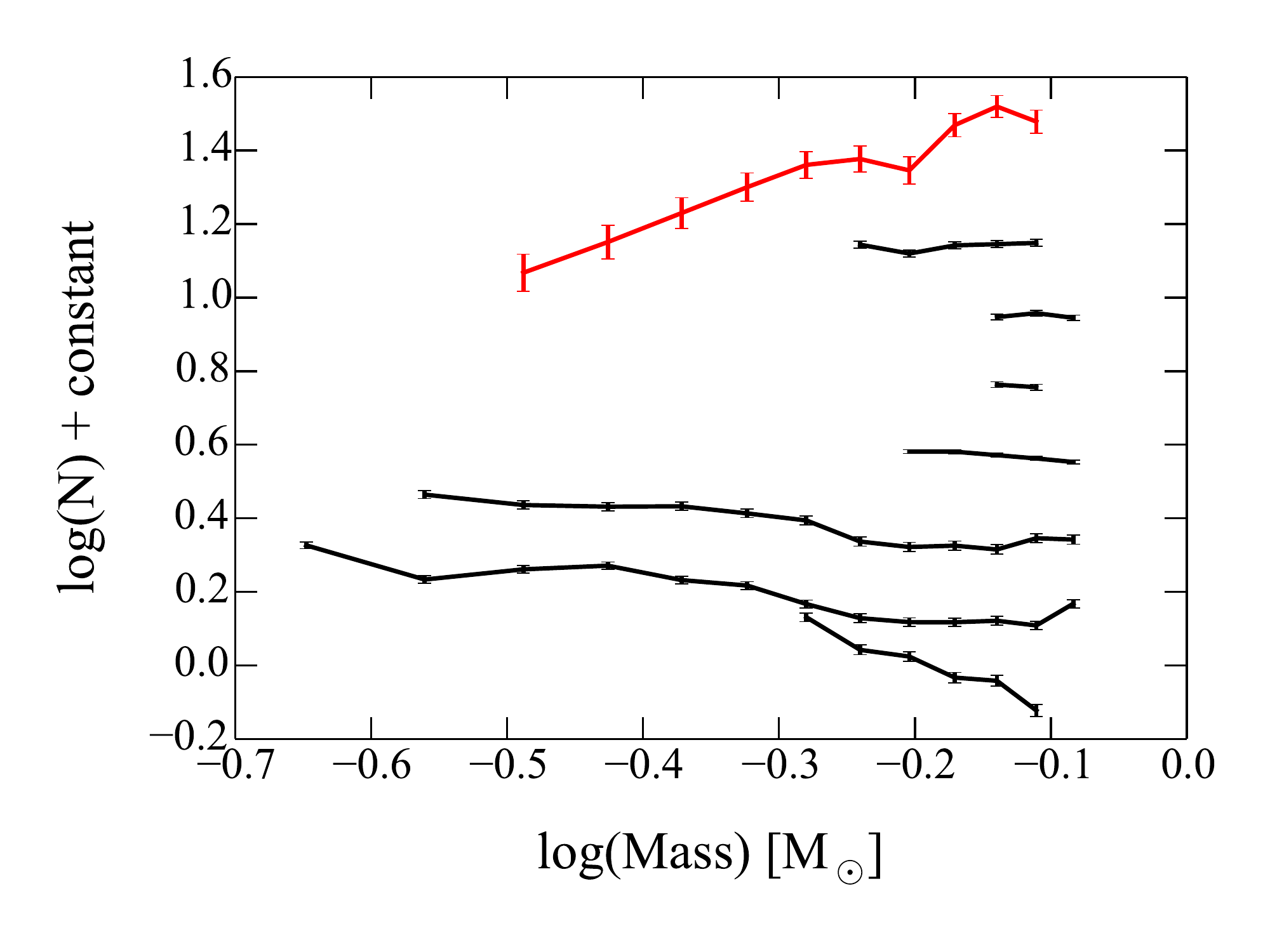}
\caption{Present day mass functions of NGC 6535 (red) and other clusters of similar age and metallicity (black) in bins of width 0.05 $M_\odot$. Error bars represent Poisson errors. Completeness corrections have been applied, but are less than a factor of two throughout. Each mass function has been shifted in the y-direction by an arbitrary constant. The clusters are, from top to bottom, NGC 6535, NGC 6981, NGC 1261, NGC 5904, NGC 288, NGC 3201, NGC 6934 and Arp 2. Unlike the other clusters, NGC 6535 has a PDMF with a positive slope.\newline}
\label{MFcomp}
\end{figure}

We present the PDMFs for these clusters in Figure \ref{MFcomp}. The clear result is that unlike those of the comparison clusters, the PDMF of NGC 6535 has a positive slope.

\subsection{Other Bottom-Light Clusters}
\label{Bottom-light}

\begin{figure}[htbp]
\centering
\includegraphics[scale=0.45]{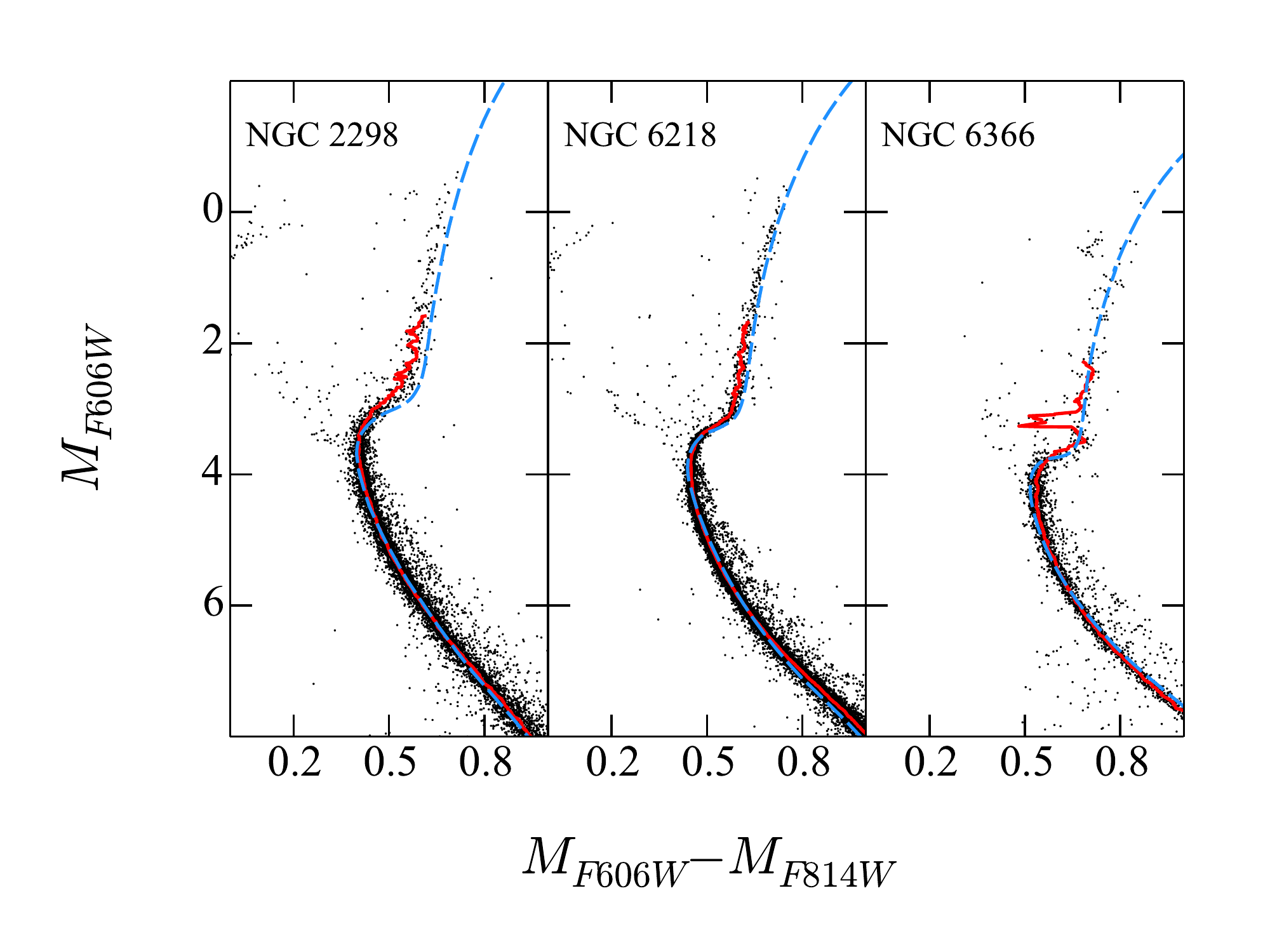}
\caption{Color-magnitude diagrams of the bottom-light clusters NGC 2298, NGC 6218 and NGC 6366. The ridgeline (solid red line) and isochrone (dashed blue line) are also plotted for each cluster. The distances and extinctions are calculated by a least squares fit. See text for details.}
\label{CMDbl}
\end{figure}

\begin{figure}[htbp]
\centering
\includegraphics[scale=0.45]{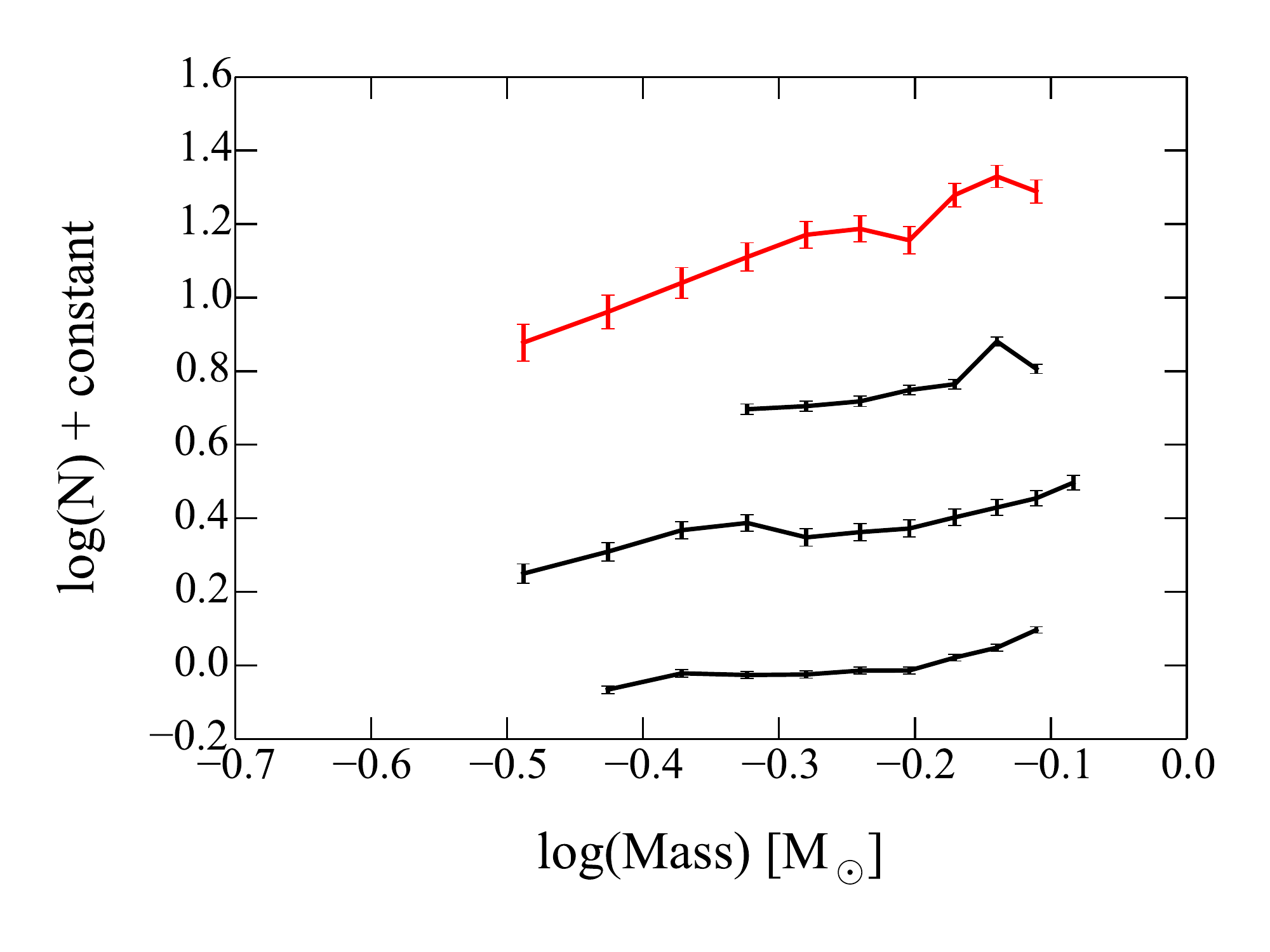}
\caption{Present day mass functions of NGC 6535 (red) and other clusters with bottom-light mass functions (black) in bins of width 0.05 $M_\odot$. Error bars represent Poisson errors. Completeness corrections have been applied, but are less than a factor of two throughout. Each mass function has been shifted in the y-direction by an arbitrary constant. The clusters are, from top to bottom, NGC 6535, NGC 2298, NGC 6366 and NGC 6218. The PDMF of NGC 6535 is qualitatively similar to the PDMFs of these bottom-light clusters.\newline}
\label{MFbl}
\end{figure}

The PDMF of NGC 6535 is unusual, but not without precedent. The globular clusters NGC 6712 \citep{NGC6712}, Pal 5 \citep{Pal5}, NGC 6218 \citep{NGC6218}, NGC 2298 \citep{NGC2298} and NGC 6366 \citep{NGC6366} have all been found to have bottom-light stellar mass functions. Of these, NGC 6218, NGC 6366, and NGC 6712 are within 5 kpc of the galactic center, as is NGC 6535 \citep[2010 edition]{Harris}, suggesting that gravitational interactions may cause clusters in this region to lose many of their low-mass stars \citep{NGC6366}. Three of these clusters, NGC 2298, NGC 6218 and NGC 6366 are included in the ACS Survey of Galactic Globular Clusters. We perform our analysis on these clusters as in \S\ref{PDMF}. We plot the CMDs of these clusters in Figure \ref{CMDbl} and compare the PDMFs of these clusters with that of NGC 6535 in Figure \ref{MFbl}. We find that like NGC 6535, these bottom-light clusters have PDMFs with positive power-law slopes. We present the parameters used for the fits in Table \ref{BLParameters}.


\afterpage{%
\begin{turnpage}
\begin{deluxetable*}{lccccccccccccc}
\tabletypesize{\scriptsize}
\tablewidth{0pt}
\tablecaption{Bottom-Light Cluster Parameters}
\tablehead{
\colhead{Cluster} &
\colhead{[Fe/H]} &
\colhead{Age} &
\colhead{$r_e$} &
\colhead{$r_{max}$} &
\colhead{$m_{F606W,max}$} &
\colhead{$D_{I}$} &
\colhead{$D_{H}$} &
\colhead{$D_{E}$} &
\colhead{$E(B-V)_{I}$} &
\colhead{$E(B-V)_{H}$} &
\colhead{$E(B-V)_{E}$} &
\colhead{ACS} &
\colhead{$\Upsilon_{dyn}/ \Upsilon_{pop}$} \\ &&
\colhead{(Gyr)}&\colhead{$(\ \prime \ )$}&\colhead{$(\ \prime \ )$}&&\colhead{(kpc)}&\colhead{(kpc)}&\colhead{(kpc)}&&&& \\
}
\startdata
NGC 2298 & $-1.92$ & 12.7 & 0.98 & 0.98 & 24.2 & 10.2 & 10.8 & 10.1 & 0.25 & 0.14 & 0.22 & Yes & No \\
NGC 6218 & $-1.37$ & 12.7 & 1.77 & 1.67 & 23.7 & 5.1 & 4.8 & 5.5 & 0.22 & 0.19 & 0.23 & Yes & Yes \\
NGC 6366 & $-0.59$ & 13.3 & 2.92 & 1.67 & 25.5 & 3.4 & 3.5 & 3.4 & 0.72 & 0.71 & 0.83 & Yes & Yes \\
NGC 6712 & $-1.02$ & ... & 1.33 & ... & ... & ... & ... & ... & ... & ... & ... & No & Yes \\
\enddata
\tablecomments{Metallicities and half-light radii from \protect\citet[2010 edition]{Harris}. Ages from \protect\citet{Ages}. See text for definitions of the various distance and extinction values. The last two columns in the table indicate whether the cluster is included in the ACS Survey of Galactic Globular Clusters and the study of \protect\citet{McLvdM}.}
\label{BLParameters}
\end{deluxetable*}  
\end{turnpage}
}

\subsection{Fainter Stars}
\label{Faint}

Below our self-imposed magnitude cutoff in some of the clusters, including NGC 6535, the mass function seems to increase dramatically. This is interesting because of the potential of these faint sources to explain this cluster's elevated dynamical mass-to-light ratio. A visual inspection of the images of NGC 6535 shows that some of these sources, particularly those with very blue colors, appear to be spurious, which is the original motivation for our magnitude cutoff. However, we now explore the nature of that population in more detail. A simple color selection that removes the faint blue sources eliminates most of the sources that do not have visual counterparts, but the upturn in the mass function remains to some degree.

To estimate the number of faint sources in the NGC 6535 images in a more quantitative, well-justified way, we match sources between the \citet{Catalogs} catalog and those found by running SExtractor on the images. We find that a slight upturn in the mass function for $M<0.3M_{\odot}$ remains, but we have not corrected for contamination by background galaxies and Galactic stars. Even so, the dramatic rise seen in the \citet{Catalogs} sources is removed, though this could be due at least in part to incompleteness in the SExtractor sources. Although a high number of very faint stars could have helped resolve the discrepancy between the observed mass function of NGC 6535 and its dynamical mass-to-light ratio, we find no such rise for $M<0.3M_{\odot}$.

\section{DISCUSSION}
\label{Discussion}

\begin{figure}[t]
\centering
\includegraphics[scale=0.45]{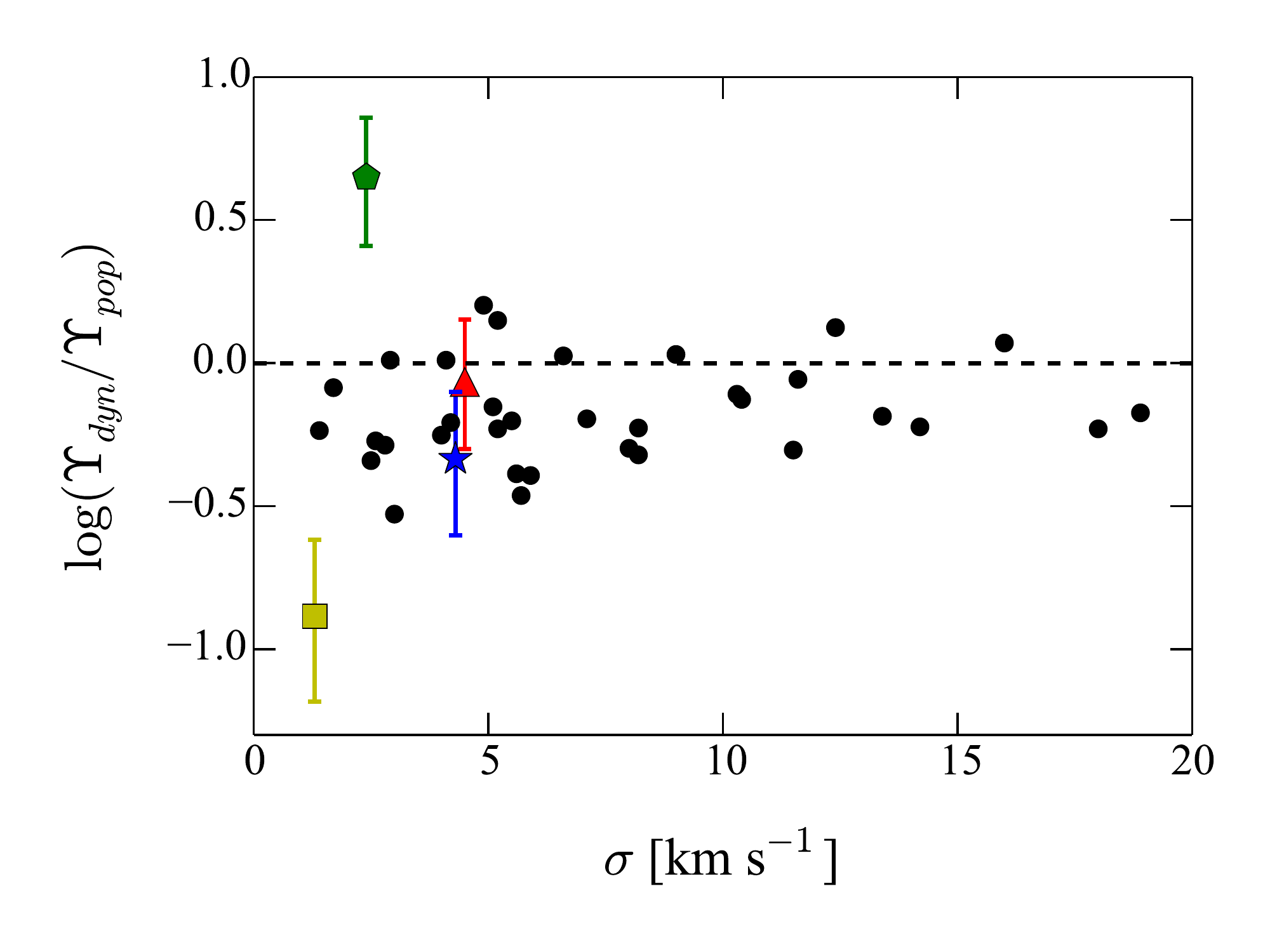}
\caption{The ratio of the dynamical ($\Upsilon_{dyn}$) and stellar population ($\Upsilon_{pop}$) mass-to-light ratios of Milky Way clusters vs. their velocity dispersions. All values are from Table 13 of \protect\citet{McLvdM}. Clusters with bottom-light mass functions are plotted as a red triangle (NGC 6218), a yellow square (NGC 6366), a green pentagon (NGC 6535) and a blue star (NGC 6712). The dynamical mass-to-light ratios neither systematically reflect the bottom-light PDMFs, nor are they universally inflated. If these clusters have been dynamically affected, the effect on the mass-to-light ratio is not straightforward.}
\label{ML}
\end{figure}

Taking into account evaporation rates and tidal shocks, \citet{Lifetimes} predict the lifetimes of globular clusters. As pointed out by \citet{NGC6218}, different studies of destruction rates are not always consistent and depend on parameters that may not be well-constrained. Nevertheless, we compare the average of the predicted lifetimes of the comparison clusters from \S\ref{Comparison}, 35.5 Gyr, to the average of the predicted lifetimes of the bottom-light clusters from \S\ref{Bottom-light}, 11.6 Gyr. It is not a large interpretative leap to suggest that the dynamics of the bottom-light clusters may be grossly affected by external processes. If these dynamical effects are the sole cause of the high dynamical mass-to-light ratio of NGC 6535,  the other bottom-light clusters may have similarly high dynamical mass-to-light ratios. Of the bottom-light clusters, NGC 6218, NGC 6366, NGC 6535 and NGC 6712 are included in Table 13 of the study of \citet{McLvdM}, which provides mass-to-light ratios calculated using the dynamical masses corresponding to a Wilson model ($\Upsilon_{dyn}$) and mass-to-light ratios calculated by stellar population models ($\Upsilon_{pop}$). In Figure \ref{ML} we plot the ratio $\Upsilon_{dyn}/\Upsilon_{pop}$ vs. the velocity disperson for all of the included Milky Way clusters, highlighting the clusters with bottom-light PDMFs. NGC 6535 is the only bottom-light cluster in this sample for which the ratio $\Upsilon_{dyn}/ \Upsilon_{pop}$ exceeds 1, and it therefore does not always follow that such clusters will have their velocity dispersion measurements and subsequent dynamical mass estimates artificially inflated. We conclude that NGC 6535 has a bottom-light PDMF similar to those of other globular clusters that may have experienced significant dynamical evolution, we may even conclude that it has experienced strong external influences that have led to the unusual PDMF, but we cannot yet conclude that those effects have artificially inflated the dynamical mass estimate.

Another cause for concern in the analysis of NGC 6535 is the large velocity anisotropy, $\sigma_t/\sigma_r = 0.79$, measured in a recent proper motion study of 22 clusters \citep{HSTPROMO}. This anisotropy is the largest measured in their sample and could, in principle, invalidate a generalized mass estimator. However, the exploration by \cite{Walker} of the mass estimator we use and that of \citet{Wolf} of related estimators suggest that this level of anisotropy should not result in sufficiently large errors to reconcile the dynamical mass-to-light ratio and the PDMF. They find uncertainties in the mass estimators on the order of a few tens of percent for a wide range of systems. Furthermore, the proper motion measurements of the velocity dispersion presented by \cite{HSTPROMO} confirm the spectral, line-of-sight velocity measurement \citep{Z14} and demonstrate that the spectral measurement was not distorted by contamination from binary stars, which would not have a comparable impact on proper motions. We conclude that the large velocity anisotropy is cause to be concerned, but does not directly translate to the large mass discrepancy between the dynamical mass estimator and a bottom-light PDMF.

If one concludes that the $\Upsilon_{*,10}$ estimate is accurate, then the cause of such a large value cannot be low-luminosity stars, barring bizarre behavior of the mass function at even lower masses than observed here. Instead, NGC 6535 could contain a large amount of mass in stellar remnants or dark matter. The overabundance of stellar remnants in this cluster relative to others could point to an unusually top-heavy IMF or again to strong dynamical evolution. The presence of dark matter in NGC 6535 would require an explanation for why it is not present in other, similar clusters. Unfortunately, our data do not provide additional constraints on these possibilities.

\begin{figure}[htbp]
\centering
\includegraphics[scale=0.45]{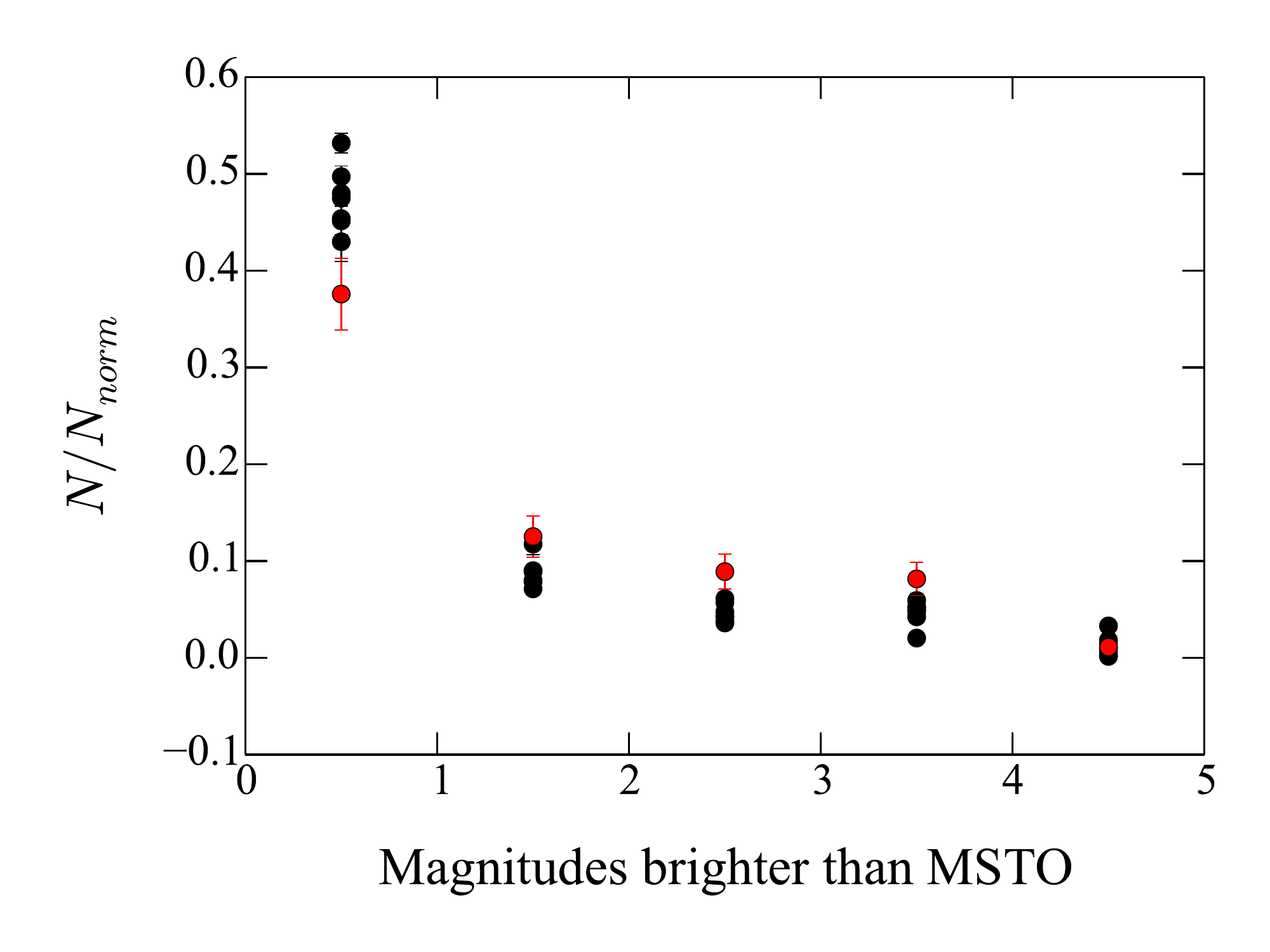}
\caption{The ratio of the completeness-corrected number of stars in magnitude-wide bins above the main sequence turnoff to the number of stars in a bin defined by $M_{F606W,MSTO}<M_{F606W}<M_{F606W,MSTO}+1$. The rightmost bin includes all stars brighter than $M_{F606W,MSTO}-4$. NGC 6535 is shown in red while other clusters of similar age are shown in black. NGC 6535 does not have an atypically low fraction of giant stars, so a lack of luminous stars does not explain its elevated mass-to-light ratio.}
\label{giants}
\end{figure}

In interpreting the high value of $\Upsilon_{*,10}$, we have focused on accounting for unseen mass, but not yet for the possibility of missing luminosity. Perhaps NGC 6535 is underluminous because it is missing a fraction of highly luminous stars. To explore this possibility, in Figure \ref{giants} we compare the completeness-corrected number of stars in magnitude-wide bins along the giant branch and horizontal branch among the clusters of similar age. We normalize between clusters by taking ratios of the number of stars in each bin to the number of stars in a bin defined by $M_{F606W,MSTO}<M_{F606W}<M_{F606W,MSTO}+1$. We find that NGC 6535 is not underpopulated in luminous stars.

Based on the open questions regarding this cluster and its unusual PDMF, we conclude that it is best to exclude NGC 6535 from current discussions regarding the possibility of IMF variations among clusters. Because this cluster was the only old cluster and the only Milky Way cluster in the high $\Upsilon_{*,10}$ population, its removal reopens questions about whether the differences in $\Upsilon_{*,10}$ are due to age or host galaxy and whether there are unknown age-dependent systematic errors. Some of these concerns may be alleviated with observations of another old cluster, NGC 2257, that appears, on the basis of less precise velocity dispersion measurements \citep{McLvdM}, to have a high value of $\Upsilon_{*,10}$.

The difficulties experienced with this one high $\Upsilon_{*,10}$ cluster may cause one to question the results for the other high $\Upsilon_{*,10}$ clusters. However, the other clusters in that set are less likely to have experienced significant dynamical evolution because they are younger and reside in lower density environments outside of the Milky Way. Nevertheless, measuring their PDMFs is critical to determining whether the high $\Upsilon_{*,10}$ values are truly due to a bottom-heavy IMF, rather than, for example, stellar remnants or dark matter. Unfortunately, comparable observations for these clusters are more difficult than for NGC 6535 because of the larger ($>8\times$) distances. IMF studies based on resolved stars have been done at these distances (e.g. \citealt{Kalirai}) but they require extremely deep Hubble Space Telescope observations.

\section{SUMMARY}
\label{Summary}

We measure the PDMF of NGC 6535, the only nearby cluster with a large dynamically measured mass-to-light ratio \citep{Z14}, using published HST data to determine whether the large mass-to-light ratio is due to a bottom-heavy IMF. We compare the PDMF to those of other globular clusters of similar age and metallicity to minimize the potential for discrepancies due to the modeling of the stellar populations. We find that the PDMF of NGC 6535 is unusually bottom-light, which exacerbates the discrepancy between the mass function and the measured mass-to-light ratio, and conclude that the large apparent dynamical mass-to-light ratio does not indicate a bottom-heavy IMF in this cluster.

To explore this discrepancy further, we compare the PDMF of NGC 6535 to those of three other bottom-light clusters. We find that the PDMFs are quite similar, suggesting that these clusters may have experienced similar histories. \cite{NGC6366} suggested, on the basis of the proximity of some of these clusters to the galactic bulge, that tidal stripping of low-mass stars, which is aided by mass segregation, could lead to accelerated loss of low-mass stars. NGC 6535 is also near the galactic bulge, and therefore the PDMF we present supports this scenario. This association of NGC 6535 with strong external effects leads to a natural supposition that its internal kinematics are sufficiently distorted by this interaction to affect the mass estimate. However, the mass estimates of other bottom-light clusters are not artificially inflated, thereby demonstrating that whatever is happening to these clusters does not necessarily lead to inflated mass estimates. The relatively high velocity anisotropy of NGC 6535 \citep{HSTPROMO} is also a cause for concern, although investigations by \cite{Walker} and \citet{Wolf} suggest that this should not affect the mass estimation enough to explain the discrepancy between the dynamical mass-to-light ratio and the PDMF.

All of this leads to a number of unresolved open questions regarding NGC 6535, and therefore this cluster is not suitable for exploring whether the IMF is universal among stellar clusters. The dynamical evolution experienced by NGC 6535 is unlikely to afflict the other known high mass-to-light ratio clusters because they are not near the galactic bulge. The disqualification of NGC 6535 from the sample is unfortunate in that it was the sole old and sole Milky Way cluster in the set of apparently high mass-to-light ratio clusters. Its removal therefore reintroduces the possibility that the $\Upsilon_{*,10}$ differences are caused by age or host galaxy. Observations of the remaining clusters in this population using methods that do not rely on dynamical mass calculations are needed to determine whether the high dynamically derived $\Upsilon_{*,10}$ values are due to a bottom-heavy IMF, other unseen mass, or simply reflect some yet unappreciated systematic error.

\acknowledgments{This work is based on observations made with the NASA/ESA Hubble Space Telescope, and obtained from the Hubble Legacy Archive, which is a collaboration between the Space Telescope Science Institute (STScI/NASA), the Space Telescope European Coordinating Facility (ST-ECF/ESA) and the Canadian Astronomy Data Centre (CADC/NRC/CSA). We thank Ata Sarajedini for assistance in understanding the artificial star catalogs from the ACS Survey of Galactic Globular Clusters. We thank the anonymous referee for providing useful comments.}

\bibliography{references}
\bibliographystyle{apj}

\end{document}